\documentstyle[twoside,fleqn,espcrc2,epsf]{article}

\newcommand{\AmS}{{\protect\the\textfont2
  A\kern-.1667em\lower.5ex\hbox{M}\kern-.125emS}}
 
\title{
Genetic Algorithm for SU(2) Gauge Theory on a 2-Dimensional Lattice
}

\author{A.Yamaguchi \address{Particle Physics Laboratory, Department 
of Physics,
        Ochanomizu University,Tokyo, Japan}}%

\begin{document}

\begin{abstract}
A hybrid algorithm is proposed for pure SU(N)
lattice gauge theory based on Genetic Algorithms (GA)s and 
the Metropolis method.  We apply the hybrid GA to pure 
SU(2) gauge theory on a 2-dimensional lattice and
find the action per plaquette and Wilson loops being
consistent with those given by the Metropolis
and Heatbath methods. The thermalization
of this newly proposed Hybrid GA is quite
faster than in the Metropolis algorithm. 
\end{abstract}

\vspace{-1cm}
\maketitle

\section{Introduction}
The study of Genetic Algorithms based upon the theory of evolution
originates with John Holland~\cite{Holland} in the mid-1970s.
Applications in various fields have been proposed.
In generic GAs, potential solutions are represented as symbolic strings and
operators are defined on the analogy of selection and genetic mechanisms
in Nature.
A crossover operator of exchanging genes between
individuals leads to the special feature of GAs in which discrete
points in searching space are treated at once,
while normal Metropolis (MP) or heat bath (HB) algorithms treat only one point.
We proposed a hybrid GA and applied it to pure SU(2) gauge theory 
in 2 dimensions.
Physical values obtained by our hybrid GAs, such as the action per plaquette
 and Wilson loops at each
$\beta$ are consistent with those given by HBs.
Our results show that the distinctive features of this new GAs
leads to be the fast thermalization.
The short calculation time are accomplished by the special encoding
configurations on a lattice.

\begin{figure*}[tb]
\begin{center}
\begin{tabular}{ccc}
\hspace{0.3cm} $\beta = 0.5$ &
\hspace{0.3cm} $\beta = 2.0$ &
\hspace{0.3cm} $\beta = 8.0$ \\
\epsfxsize=4.5cm\epsffile{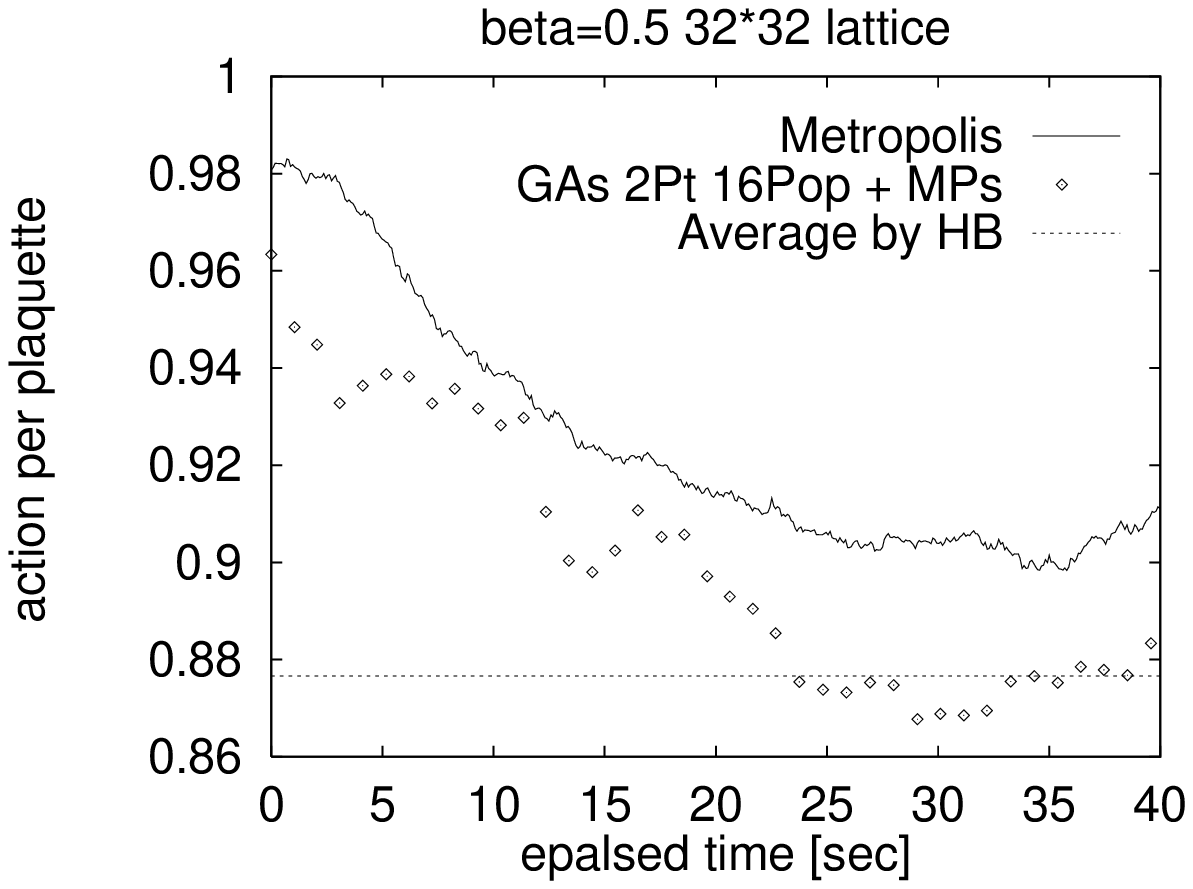} &
\epsfxsize=4.5cm\epsffile{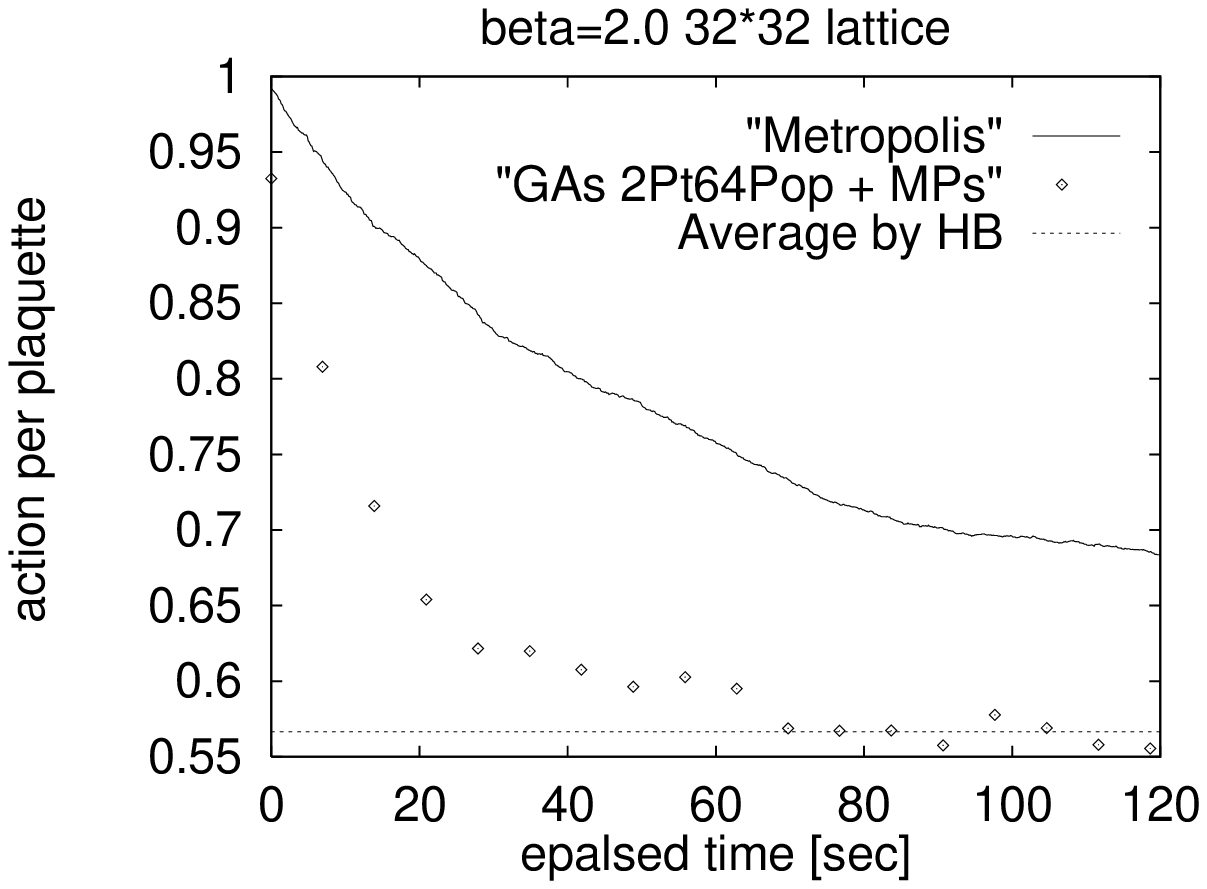}  &
\epsfxsize=4.5cm\epsffile{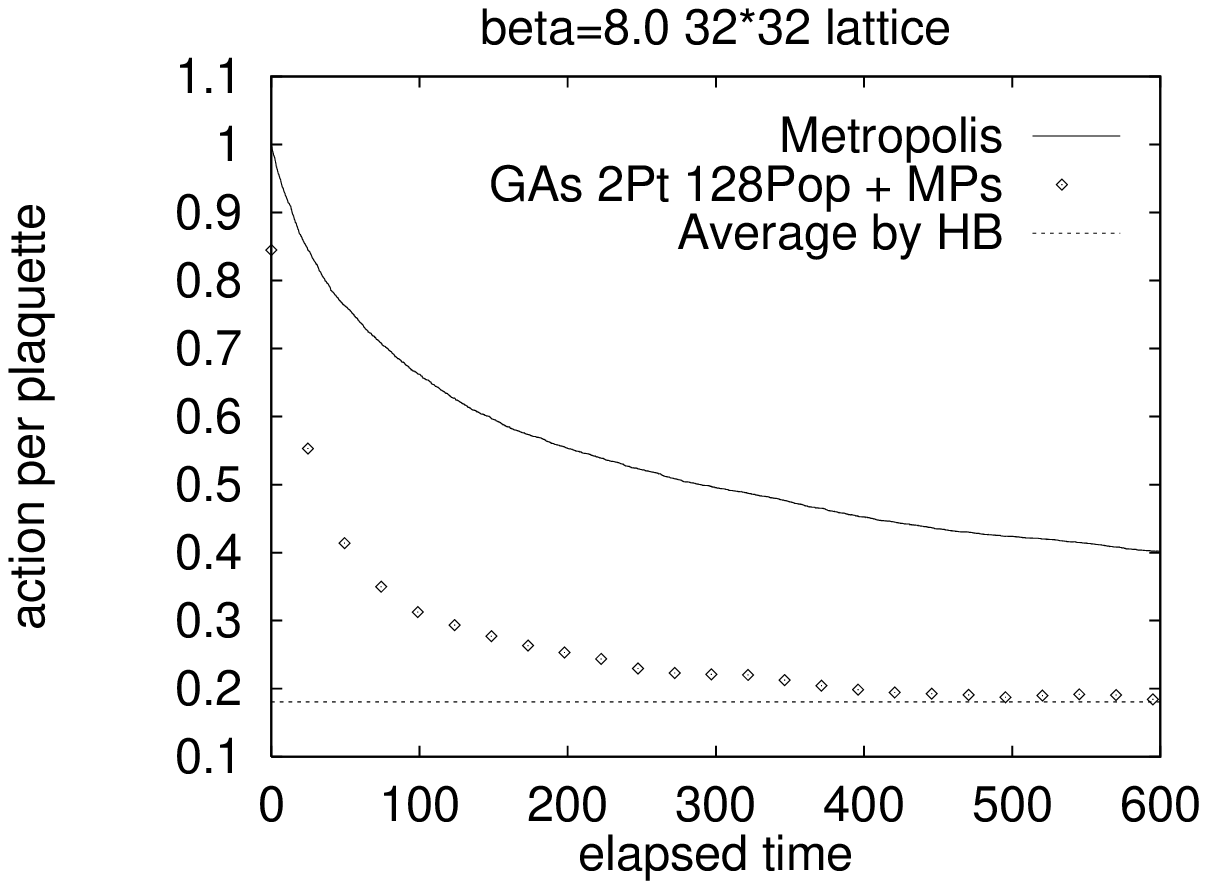} \\
\end{tabular}
\end{center}
\vspace{-1.2cm}
\begin{center}
\caption{ \label{fig1}
Thermalization of the action per plaquette on $32 \cdot 32$ lattice}
\end{center}
\vspace{-1cm}
\end{figure*}
\section{
Genetic Algorithms for SU(N) Lattice gauge theory}
 
In the language of GAs, a whole lattice corresponds to an individual, and
the information on a field configuration is
encoded in 
bit patterns which are treated as chromosome.
The evaluation (fitness) function is the usual lattice action
of SU(N) lattice gauge theory,
\begin{equation}
\displaystyle{
  S[U] = \Sigma_p \left( 1 - \frac{1}{N}ReTrU_p \right),
             }
\end{equation}
where $U_p$ is the element of SU(N) defined on a plaquette $p$ and $\beta$ is
a coupling constant multiplying $S$ in the Boltzmann factor.

We would like to call our newly proposed algorithm ``Hybrid Genetic Algorithm''
(HGA), because it mixes a population update step of GAs with a normal
Metropolis step applied to all members of the population.
It combines 
heuristic methods for local search with GAs for global search
with their respective advantages.
In our algorithm, the GA starting from the $t^{\mathrm{th}}$
generation is followed by the Metropolis step, leading to the next
$(t+1)^{\mathrm{th}}$ generation, namely
\begin{eqnarray}
\vspace{-0.5cm}
P_t \stackrel{selection}{\longrightarrow}
& P_{gene~pool} & \stackrel{recombination}{\longrightarrow} \\
\nonumber 
&  P_{gene~pool}^{\prime} &\stackrel{Metropolis}{\longrightarrow}
P_{t+1},
\vspace{-0.5cm}
\end{eqnarray}
where $P_t$ denotes the population at $t^{\mathrm{th}}$ generation  
and $P_{gene~pool}$ represents a number of $2N_{pop}$ genomes selected by
Stochastic Universal Sampling which
ensures unbiased selection~\cite{Baker}.
In this scheme, among all genomes of $P_t$ laid on a pie chart 
with an area proportional to their
fitness, $2N_{pop}$ members are picked up simultaneously into a gene pool 
by a single spin of a roulette wheel with $2~N_{pop}$
equally-spaced pointers.  
Two genomes taken from the gene pool are recombined by 
the crossover operator to give two offsprings (children).
The children are subject to a mutation step which consists of bitwise
random flips described by a (small) mutation rate. 
To establish thermal equilibrium, we set up an accept/reject criterion
in the next stage to fulfill the detailed balance condition: 
one does not replace {\it automatically} old individual of the previous 
generation by new one but 
the better parent is replaced by the better child when the following condition
is satisfied:
\begin{equation}
 min\{1, exp^{-\beta*(S_{child} - S_{parent})} \} > \xi,
\end{equation}
where $\xi$ is a uniform random number\cite{gene0}.
The population $P_{t+1}$ is then the result of one normal Metropolis step.

\section{Thermalization}

In order to compare the thermalization rate with MP, 
a $2$-point crossover scheme is adopted with a crossrate $0.65$
and a mutation rate
$0.008$. The population sizes are $16$ for $\beta = 0.5$,
$64$ for $\beta =2.0$  and $128$ for $\beta = 8.0$.
Fig.~\ref{fig1}
shows the results of these runs
for three $\beta$ values.
In each figure, the horizontal dotted line shows
the averaged action per plaquette for this $\beta$ over 
the last $1000$ sweeps of $30,000$ sweeps by the Heatbath method.
Square dots show the time history corresponding 
to our HGA, the lines show the MC history of pure MP. 
The convergence to equilibrium (the target values given by HB)
for the runs of our HGA is faster in CPU time than MP. 
In particular, for large $\beta$ it is quite fast not being stuck
due to low acceptance.
The approach to equilibrium in the 
run with $\beta = 0.5$ is, however, non-monotonous due to 
hard fluctuations.
In the case of small $\beta$s,
most genomes in a population have very similar actions, 
which could have very similar genes or very different ones, an unexpected  
change of action is likely to be produced 
and accepted by the crossover operation.
To avoid strong fluctuations caused by 
bad crossing, a closer 
analysis of the correlation between individuals is needed.
Note that while MP treats just one lattice,
HGA treats a population size number of lattices at once.

\begin{figure}[hbt]
\vspace{-0.6cm}
\begin{center}
\begin{tabular}{cc}
\epsfxsize=3.3cm\epsffile{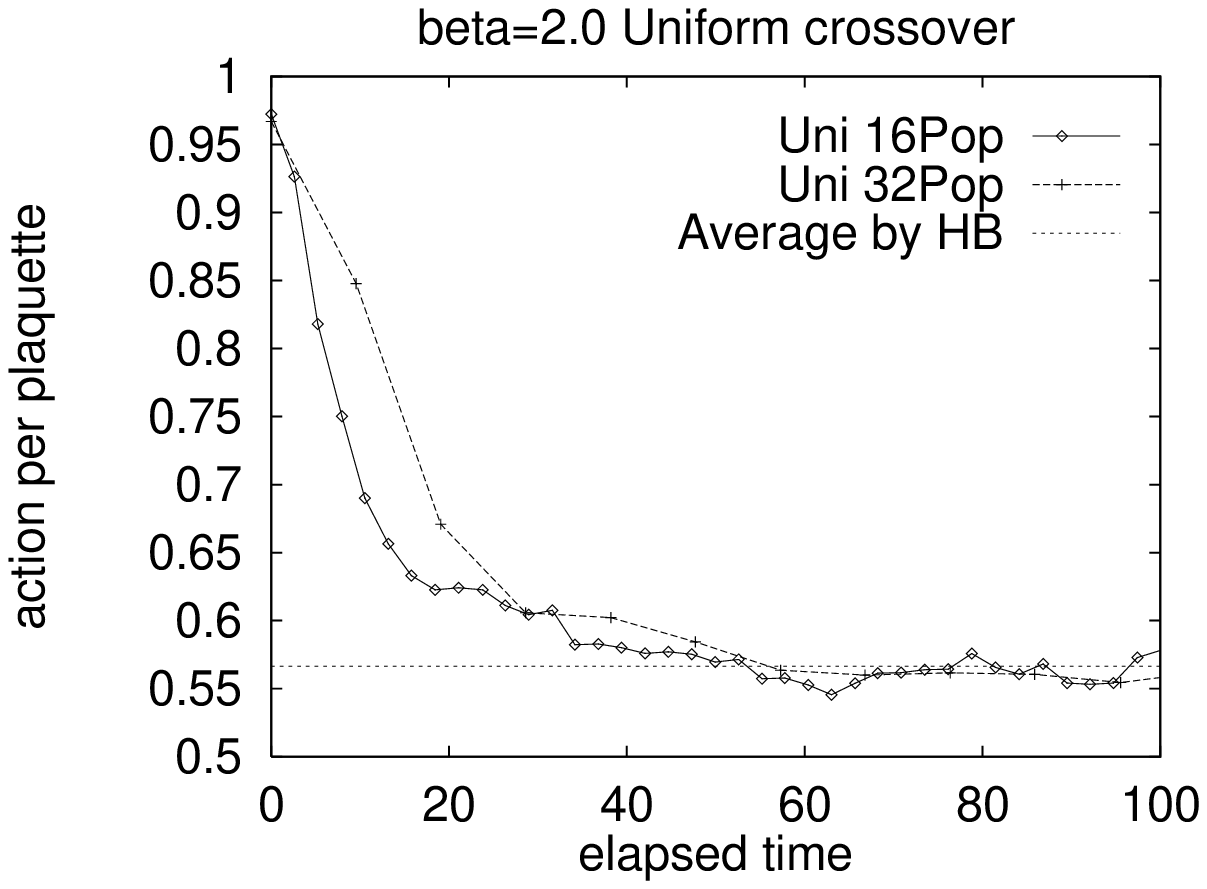}&
\epsfxsize=3.3cm\epsffile{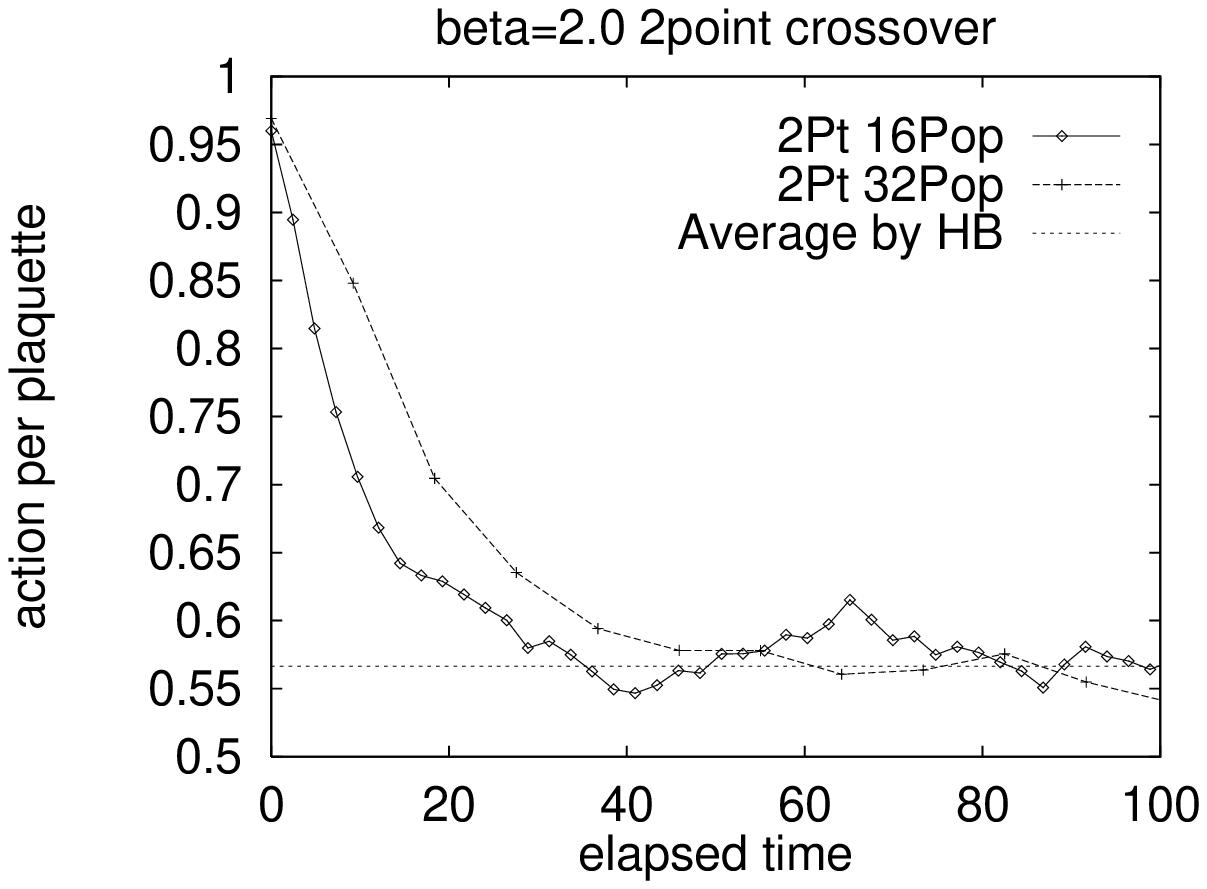}\\
\epsfxsize=3.3cm\epsffile{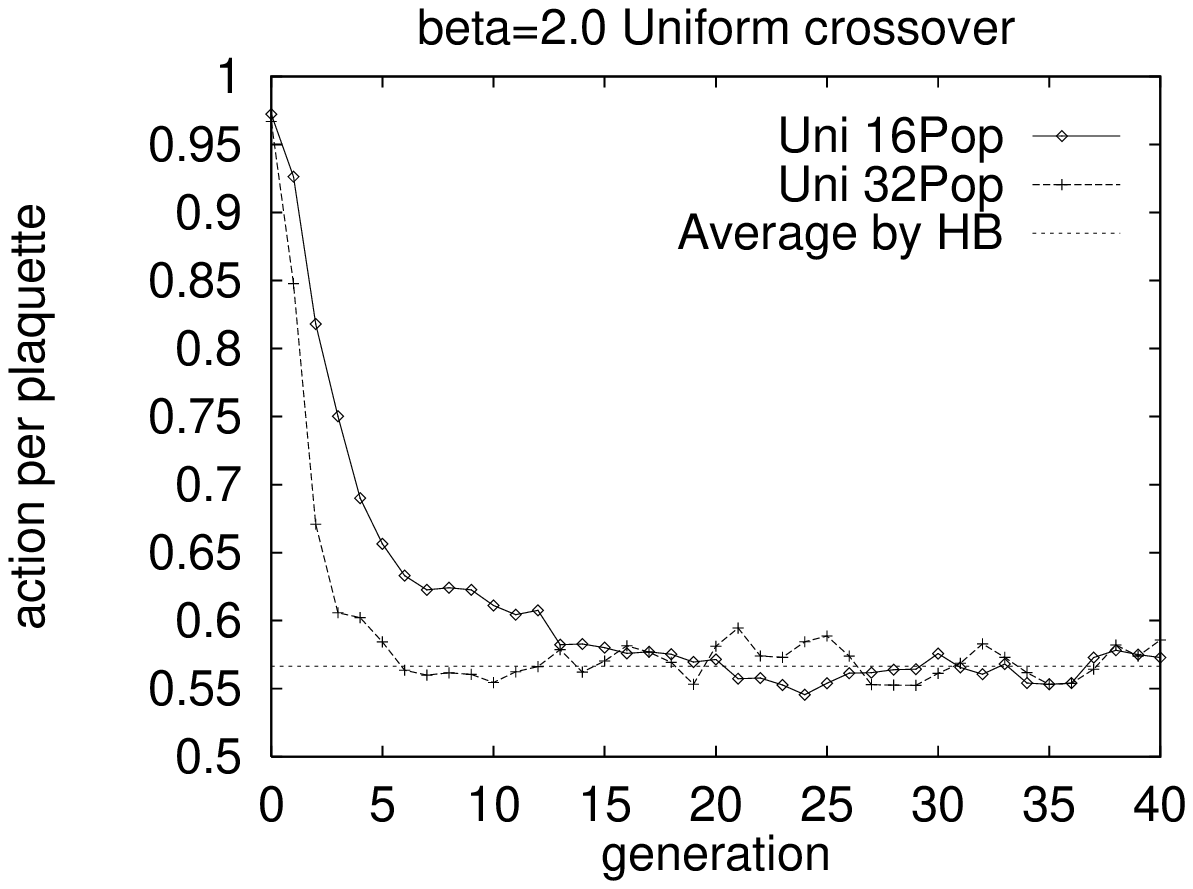}& 
\epsfxsize=3.3cm\epsffile{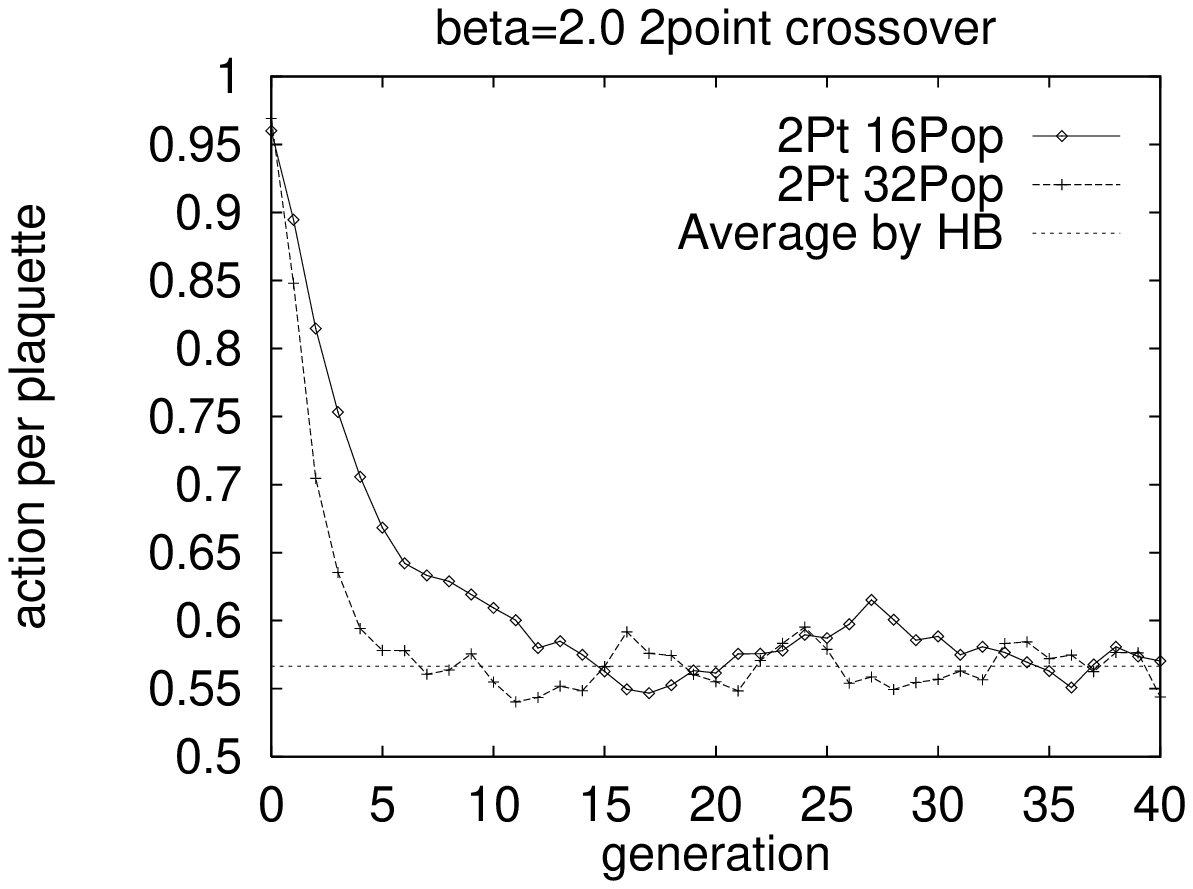}\\
\end{tabular}
\end{center}
\vspace{-1.3cm}
\begin{center}
\caption{\label{fig2}
Comparison of Recombination schemes and population sizes at $\beta=2.0$ }
\end{center}
\vspace{-1.4cm}
\end{figure}
\section{Recombination Schemes and Genetic Parameters}
We have tested
two recombination schemes uniform crossover (UC)
and $2$-point crossover (2C) with respect to their effectivity. 
UC in which every locus of a genome
is randomly occupied by a gene from one of the parents 
with a probability $p$ (or from the other with
$1-p$), leads to an increased 
diversity of offsprings. In the lattice gauge theory it 
can lead to an increase of 
action, caused by a frequent mismatch of the links. 
In $N_c$-point crossover
a new genome is created from partitions of parents genomes. 
It has less advantage for
the diversity, but it reduces the mismatch problem of UC.
To compare these schemes and to study the effect of population
size for eventual premature convergence, 
 simulations are performed with $\beta=2.0$,
with two population sizes, $32$ and $16$, with an equal 
crossrate of $0.65$.
Fig.~\ref{fig2}
shows the comparison of schemes and population sizes
for the case of $\beta=2.0$.
One sees that a large population size is useful for
rapid decrease of the action (measured by the necessary number of
generations), because it easier escapes from local
minima thanks to the richer diversity in the population. 
On the point of elapsed time, however,
the remarkable difference of convergence speed does not appear because
of the large number of procedure.
It suggests us the necessity of optimization of GA operators.
Runs with UC did not converge fast, 
because of increasing link mismatch action that tends to outweigh
the eventual decrease of bulk action.

\begin{figure}[tb!]
\begin{center}
\begin{tabular}{c}
\hspace{0.3cm} Action per Plaquette \\
\epsfxsize=6.5cm\epsffile{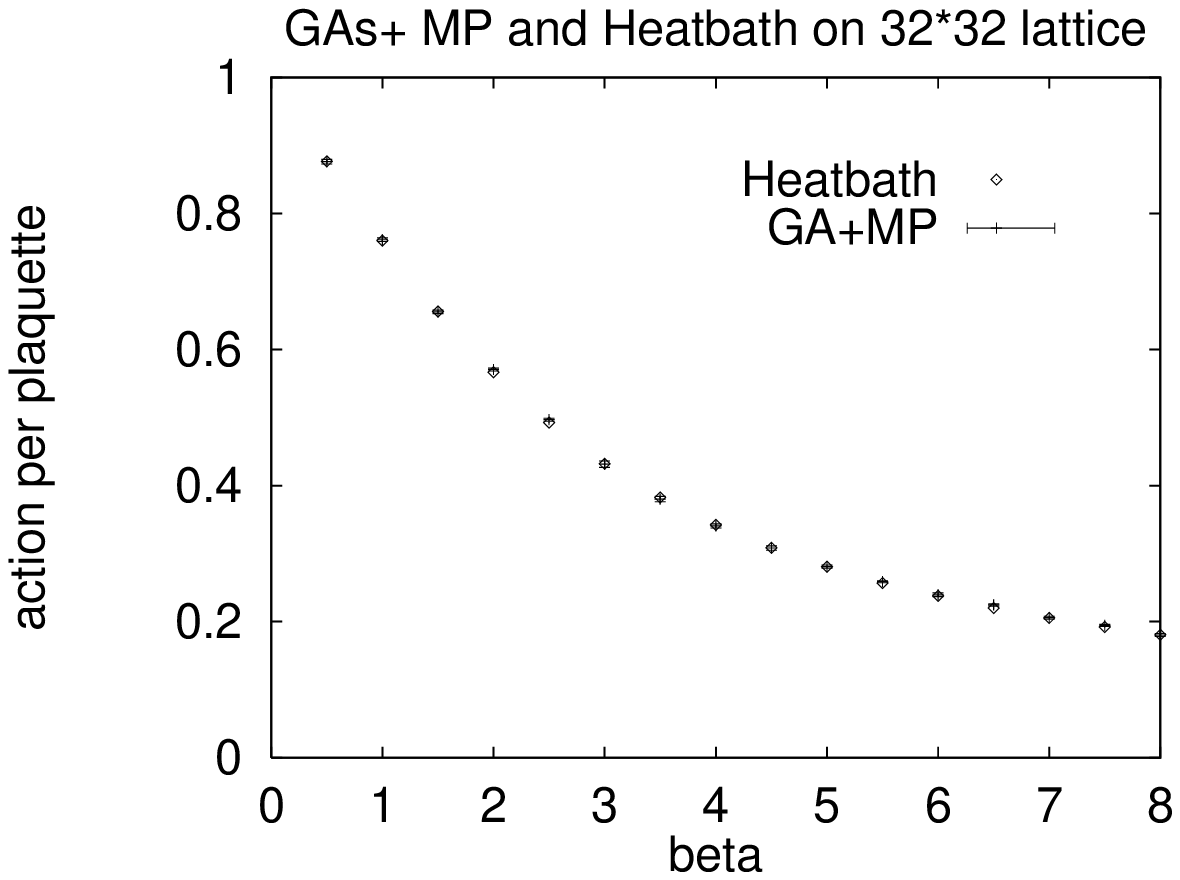} \\
\hspace{0.3cm} Wilson Loops \\
\epsfxsize=6.5cm\epsffile{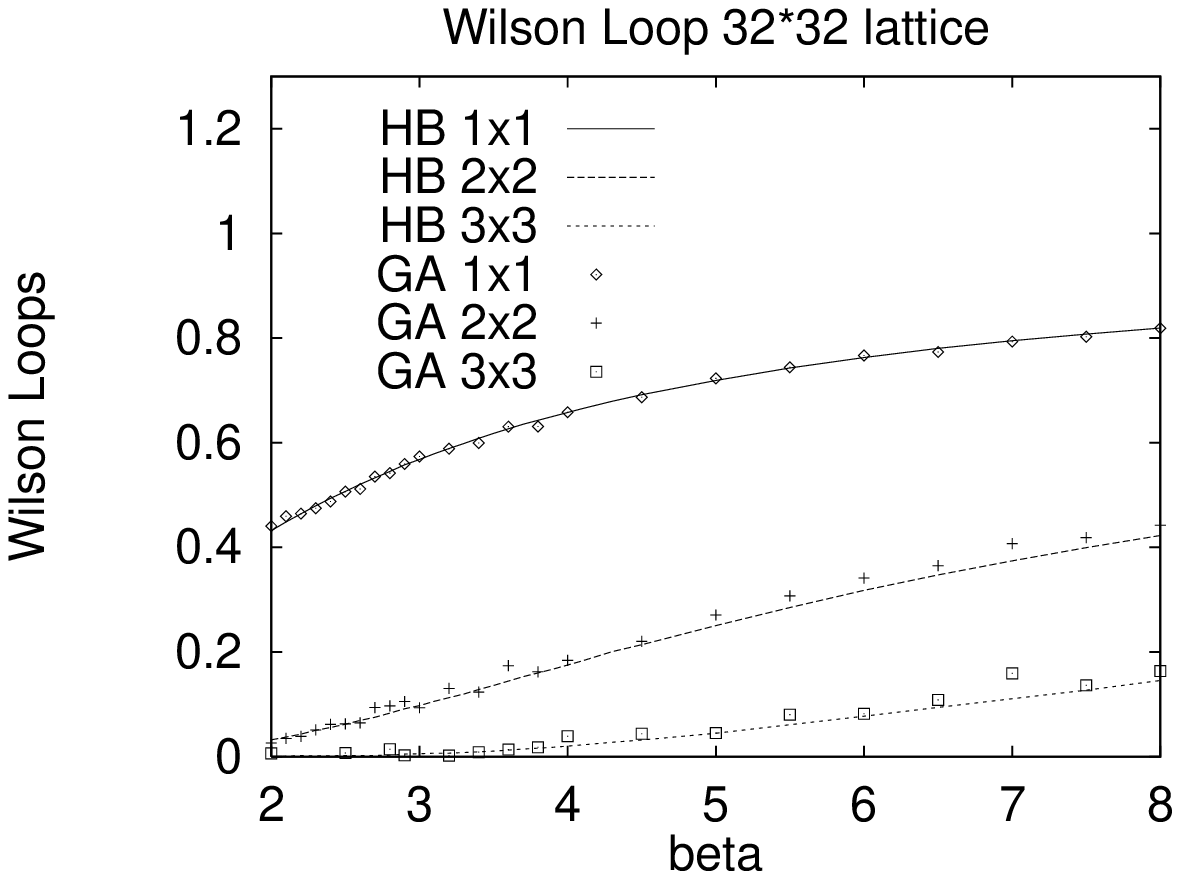} \\
\end{tabular}
\end{center}
\vspace{-1.0cm}
\caption{ \label{fig3}
 Physical values on $32 \cdot 32$ lattice}
\vspace{-1.5cm}
 \end{figure}%
\section{Conclusion}
We have designed a HGA and applied it to $2D$ $SU(2)$
lattice gauge theory and calculated the action
per plaquette and the Wilson loops. The results are shown in
Fig.~\ref{fig3}.
For small $\beta \le 2.0$ the action per plaquette obtained by HGA 
have been averaged over the last 32 generations out of 64 generations keeping 
a population size of $32$, while for large $\beta$s( $\ge 2.5$)
over 16 generations out of 32 generations with a population size of 128.
The average of Wilson loops obtained with  MP
were taken over 1000 iterations of the best genome
among the population in the $50^{\mathrm{th}}$ generation of HGA for $\beta$s 
between $2.0 \le \beta \le 4.0$
(of the best genome in the  $30^{\mathrm{th}}$ generation 
for $\beta \ge 4.5$).
Over the whole $\beta$ region the result of the physical averages 
are consistent with HB. They are obtained by HGA faster than by HB
and  normal MP,
without any further optimization of the GA part of our procedure.
We have shown the possibility and effectiveness of GA schemes 
for SU(N) lattice gauge theory.
A more detailed discussion and analysis is needed 
to demonstrate the detailed balance properties. 
There is also enough room for
further optimization of our algorithm.

\section{Acknowledgment}
I acknowledge the use of the Reproduction Plan Language, RPL2 produced
by Quadrastone Limited, and
the use of workstations of Yukawa Institute.

\end{document}